\documentclass[aps,prd,amsmath,amssymb,superscriptaddress,twocolumn,
nofootinbib,showpacs,preprintnumbers]{revtex4-1}
\usepackage{graphicx,hyperref,color,subfigure,soul}
\usepackage{cleveref,units,ctable,braket,xspace}

\hypersetup{ 
    pdfnewwindow=true,      
    colorlinks=true,       
    linkcolor=blue,          
    citecolor=blue,        
    filecolor=blue,      
    urlcolor=blue        
}  

\crefname{figure}{Fig.}{Figs.}
\crefname{equation}{Eq.}{Eqs.}
\crefname{table}{Table}{Tables}
\crefname{chapter}{Chapter}{Chapters}
\crefname{section}{Section}{Sections}
\crefname{appendix}{Appendix}{Appendices}

\newcommand{\jpsi}{\ensuremath{J/\psi}\xspace}
\newcommand{\jpsip}{\ensuremath{J/\psi\,p}\xspace}
\newcommand{\ie}{\emph{i.e.}\xspace}
\newcommand{\etal}{ \textit{et al.}\xspace}

\newcommand{\valencia}{Departamento de F\'\i sica Te\'orica and IFIC, 
Centro Mixto Universidad de Valencia-CSIC, 
Institutos de Investigaci\'on de Paterna, E-46071 Valencia, Spain}
\newcommand{\jlab}{Thomas  Jefferson  National  Accelerator  Facility, Newport  News,  VA  23606,  USA}
\newcommand{\ceem}{Center for  Exploration  of  Energy  and  Matter,  Indiana  University,  Bloomington,  IN  47403,  USA}
\newcommand{\indiana}{Physics  Department,  Indiana  University,  Bloomington,  IN  47405,  USA}
\newcommand{\unam}{Instituto de Ciencias Nucleares, Universidad Nacional Aut\'onoma de M\'exico,  
Ciudad de M\'exico 04510, Mexico}
\newcommand{\infn}{INFN Sezione di Roma, Roma, I-00185, Italy}

\begin{document}

\title{Studying the $P_c(4450)$ resonance in $\jpsi$ photoproduction off protons}

\author{A.~N.~\surname{Hiller Blin}}
\affiliation{\valencia}
\affiliation{\ceem}

\author{C.~Fern\'andez-Ram\'{\i}rez}
\affiliation{\unam}

\author{A.~Jackura}
\affiliation{\ceem}
\affiliation{\indiana}

\author{V.~Mathieu}
\affiliation{\ceem}
\affiliation{\indiana}

\author{V.~I.~\surname{Mokeev}}
\affiliation{\jlab}

\author{A.~\surname{Pilloni}}
\affiliation{\jlab}
\affiliation{\infn}

\author{A.~P.~Szczepaniak}
\affiliation{\ceem}
\affiliation{\jlab}
\affiliation{\indiana}

\preprint{JLAB-THY-16-2277}

\collaboration{Joint Physics Analysis Center}

\begin{abstract}
A resonance-like structure, the $P_c(4450)$, has recently been observed 
in the $\jpsip$ spectrum by the LHCb collaboration.  
We discuss the feasibility of detecting this structure in $\jpsi$ 
photoproduction in the CLAS12 experiment at JLab. 
We present a first estimate of the upper limit for the branching ratio of the $P_c(4450)$ to $\jpsip$. 
Our estimates, which take into account the experimental resolution
effects, predict that it will be possible to observe a sizable cross section close to the $J/\psi$  
production threshold and shed light on the $P_c(4450)$ resonance in the future photoproduction measurements.
\end{abstract}
\date{\today}
\maketitle

\section{Introduction}
Exotic hadron spectroscopy opens a new window into quark-gluon dynamics that  could shift 
the paradigm that mesons and baryons consist of $q\bar{q}$ and $qqq$ constituent quarks, respectively. 
Recent lattice QCD studies of the nucleon spectrum indicate that the excited nucleon states may exist 
with a substantial admixture of glue~\cite{Edwards:2012fx}. 
These recent predictions initiated the efforts aimed at a search for hybrid baryons in the future experiments at 
Jefferson Lab with the CLAS12  detector~\cite{clas12,Burkert:2016dxc}.  
In this paper we discuss the feasibility of using the CLAS12  detector  
in search for  exotic baryons with the quark core consisting of five constituent quarks including charm.  
This is motivated by Refs.~\cite{Wang:2015jsa,Karliner:2015voa,Kubarovsky:2015aaa}, where the authors propose to use photons 
to produce hidden-charm pentaquarks of the type that were 
reported  by the LHCb collaboration in the $\Lambda^0_b \to K^- (\jpsip)$ channel~\cite{Aaij:2015tga}.
In the LHCb data, two structures were observed, the  broader  
has a width of $205\pm18\pm86~\unit{MeV}$ and mass $4380\pm8\pm29~\unit{MeV}$, and the narrower 
has width $39\pm5\pm19~\unit{MeV}$ and mass $4449.8\pm1.7\pm2.5~\unit{MeV}$. 
The preferred spin-parity assignment of these structures is that of $J_r=3/2$ or $J_r=5/2$ and  
opposite parities.  Here we focus on the narrower structure, 
referred to as $P_c(4450)$, since we expect the broad one to be more susceptible to variations 
in the analysis model used to describe the coherent background.  
Various interpretations of these structures have been proposed. 
The possibility of a loosely-bound molecular state of charmed baryons and mesons  
was investigated in~\cite{Chen:2015loa,Chen:2015moa,Roca:2015dva,He:2015cea,Lu:2016nnt}, 
while a resonance interpretation in terms of quark degrees of freedom was proposed 
in~\cite{Maiani:2015vwa,Anisovich:2015cia,Mironov:2015ica,Lebed:2015tna}. 
The possibility that these structures are  nonresonant, for example due to the presence of nearby singularities in cross channels   
was discussed  in~\cite{Meissner:2015mza,Guo:2015umn,Liu:2015fea,Mikhasenko:2015vca} 
(for recent reviews on the exotic charmonium-like sector, see~\cite{review}).  
If the resonant nature holds, it would be the first time that a signature of a hidden-charm baryon state is found. 
It is therefore important to look for other ways to produce the 
$\jpsip$ system near threshold~\cite{Wang:2015jsa,Karliner:2015voa,Kubarovsky:2015aaa}. 
For example, if a peak in the $\jpsip$ mass spectrum appears in photoproduction, 
the nonresonant interpretation of the LHCb result  would be less likely. 
        
In this paper we make a prediction for the $\jpsi$ photoproduction cross section 
measurement for the CLAS12 experiment at JLab. 
We closely follow the arguments of~\cite{Karliner:2015voa}, 
in particular for the application of vector-meson dominance (VMD). 
To describe the baryon-resonance photoproduction we use the model of~\cite{Mokeev:2012vsa} 
that was successfully applied in the past to the analysis of    $N^*$ photo- and electroproduction in the exclusive $\pi^+\pi^-p$ channel.  
Compared to hadronic production, the exclusive
$J/\psi$  photoproduction off protons is expected to have a large $P_c(4450)$ resonant contribution relative to the background. 
Furthermore,  unlike  the LHCb case, there is no ``third" particle in the final state that the $\jpsip$ 
system could rescatter from.   
The existing photoproduction data~\cite{Camerini:1975cy,Chekanov:2002xi,Aktas:2005xu} 
mainly cover the range of photon energies above $100~\unit{GeV}$, \ie well above the possible resonance signal, 
and it can be well understood as diffractive production. 
The few  data points in the energy range of interest~\cite{Ritson:1976rj}  
have a mass resolution which is too low to identify a potential resonance signal.  
The  LHCb peak in  photoproduction is  expected in a photon energy  range where the 
diffractive cross section is rather low and one can expect a clearly visible resonance peak. 
  
The CLAS12 detector is replacing the CLAS apparatus in Hall B at JLab and was optimized for measurements
of nucleon resonances in electro- and photoproduction via decays to several exclusive meson-nucleon final states~\cite{Aznauryan:2015b}. 
The excitation of  the possible hidden charm resonance in the $\jpsip$ system requires photons with energies 
up to $11~\unit{GeV}$, and the identification of the resonance involves partial wave analysis. Therefore the 
measurement of the differential cross section and  spin-density matrix elements would be desired.  
The cross section measurement will be possible with the data from the forward tagger built into the new CLAS12 detector. 
Ultimately, if the resonance signal is found,  it would be of interest to extend 
the present study to $\jpsip$ electroproduction, to investigate its internal structure. 
The $\jpsi$-polarization information is currently not feasible with CLAS12 without muon detection capability, 
but if the signal is found it would be a good candidate for a detector upgrade.

\section{Reaction model}\label{model}

\begin{figure}
\begin{subfigure}[\ Pomeron exchange]{
\includegraphics[width=0.45\columnwidth]{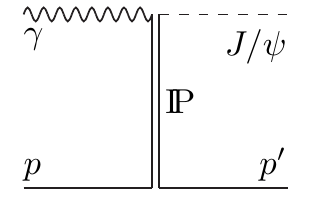}}
\label{fjpsiblob}
\end{subfigure}
\begin{subfigure}[\ Resonant contribution]{
\includegraphics[width=0.45\columnwidth]{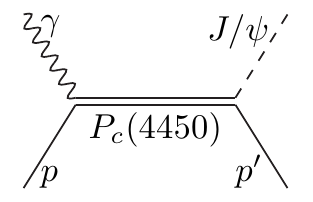}}
\label{fjpsipc}
\end{subfigure}
\caption{Dominant contributions to the $\jpsi$ photoproduction.
The nonresonant background is modeled by an effective Pomeron exchange (a)
while the resonant contribution  of the $P_c(4450)$ in the direct channel (b) 
is modeled by a Breit-Wigner amplitude.} 
\label{Fjpsiprod}
\end{figure}
%

\subsection{Resonant contribution}
The processes contributing to  $\gamma\, p \to \jpsip$ are shown in Fig.~\ref{Fjpsiprod}. 
The nonresonant background is expected to be dominated by the $t$-channel Pomeron exchange, 
and  we saturate the $s$-channel by the $P_c(4450)$ resonance. 
In the following we consider only the most favored
$J_r^P = 3/2^-$ and $5/2^+$ spin-parity  assignments for the resonance. 
We adopt the usual normalization conventions~\cite{pdg},
and express the differential cross section 
in terms of the helicity  amplitudes
$\bra{\lambda_{\psi}\lambda_{p^\prime}}T_r\ket{\lambda_\gamma \lambda_p}$, 
\begin{equation}
\label{Edsigdcos}
\frac{d\sigma}{d \cos\theta}= \frac{ 4\pi\alpha}{32 \pi  s} \frac{p_f}{p_i} \frac{1}{4} 
\sum_{\lambda_\gamma,\lambda_p, \lambda_{\psi}, \lambda_{p'}}
\left|\bra{\lambda_{\psi}\lambda_{p^\prime}} T \ket{\lambda_\gamma \lambda_p}\right|^2.
\end{equation}
Here, $p_i$ and $p_f$ are the incoming and outgoing center-of-mass frame momenta, respectively, 
$\theta$ is the center-of-mass scattering angle, and $W=\sqrt{s}$ is the invariant mass. 

Note that the electric charge $\sqrt{4\pi \alpha}$ is 
explicitly factored out from the matrix element. 
The contribution of the $P_c(4450)$ resonance is parametrized using the  
Breit-Wigner ansatz~\cite{Mokeev:2012vsa}, 
\begin{align}
\bra{\lambda_{\psi}\lambda_{p^\prime}}T_r\ket{\lambda_\gamma \lambda_p}=
 \frac
{\bra{\lambda_{\psi}\lambda_{p^\prime}}T_{\text{dec}}\ket{\lambda_R}\bra{\lambda_R}
T^\dagger_\text{em}\ket{\lambda_\gamma \lambda_p}}
{M_r^2-W^2-\mathrm{i}\Gamma_rM_r}.\label{EBWamp}
\end{align}
The numerator is given by the product of photo-excitation and hadronic decay helicity amplitudes. 
The measured width is narrow enough to be approximated with a constant, 
$\Gamma_r=(39\pm 24)\mbox{ MeV}$.  
The angular momentum conservation restricts the sum over $\lambda_R$, 
the spin projection along the beam direction in the center of mass frame, 
to $\lambda_{R}=\lambda_{\gamma}-\lambda_{p}$.   
The hadronic helicity amplitude $T_\text{dec}$, which represents the decay of the resonance 
of spin $J$ to the $\jpsip$ state, is given by 
\begin{align}
\bra{\lambda_{\psi}\lambda_{p'}}T_\text{dec}\ket{\lambda_R}=
g_{\lambda_{\psi}\lambda_{p'}} d_{\lambda_R,\lambda_{\psi}-\lambda_{p'}}^{J}(\cos\theta), 
\end{align}
where  $g_{\lambda_{\psi}\lambda_{p'}}$ are the helicity couplings between the resonance and the final state. 
There are three independent couplings with  $\lambda_{p^\prime} = \frac{1}{2}$, $\lambda_{\psi} = \pm 1,0$,  
being the other three related by parity. For simplicity, we  assume all these couplings to be equal,
\ie $g_{\lambda_{\psi}\lambda'_p}\equiv g$.
The helicity amplitudes  and  the partial decay width $\Gamma_{\psi p}$ 
are related by 
\begin{widetext}
\begin{equation}
\Gamma_{\psi p} = \mathcal{B}_{\psi p}\, \Gamma_r 
=\frac{\bar p_f}{32 \pi^2 M^2_{r}}\frac{1}{2J_{r}+1}  \sum_{\lambda_{R}}\int d\Omega \,\vert\langle 
\lambda_{\psi}\lambda_{p'} \vert T_{dec} \vert  
\lambda_{R} \rangle |^2 = \frac{\bar p_f}{8\pi M_r^2} \frac{6g^2}{2J_r + 1} ,
\label{ampwidth}
\end{equation} 
\end{widetext}
with $\mathcal{B}_{\psi p}$ being the branching ratio of $P_c \to \jpsip$
and $\bar p_f$ the momentum $p_f$ evaluated at the resonance peak.
We assume that the $P_c(4450)$ decay is dominated by the lowest partial wave, 
with angular momentum $\ell=0$ for $J_r^P=3/2^-$ and $\ell=1$ for $J_r^P=5/2^+$.
We recall that the following near-threshold behavior of the helicity amplitudes holds:
$g \propto  p_f^\ell$.

The helicity matrix elements  of $T_\text{em}$  are usually parametrized 
in terms of two independent coupling constants, $A_{1/2}$ and $A_{3/2}$, 
which are  related to the matrix elements
with $\lambda_R=1/2, 3/2$, respectively. The other two helicities $-1/2$ and $-3/2$ are constrained 
by parity.  Using the standard normalization convention, in which the
helicity couplings $A_{\lambda_{R}}$ have units of $\text{GeV}^{-1/2}$ and are proportional to the unit electromagnetic charge, 
\begin{equation}
\label{a12a32sec}
\langle \lambda_{\gamma}\lambda_{p} \left| T_\text{em} \right| \lambda_{R} \rangle  
=\frac{W}{M_{r}}\sqrt{\frac{8M_{N}M_{r} \bar p_{i}}{4\pi\alpha}}
\sqrt{\frac{\bar p_i}{p_i}}A_{1/2,3/2},
\end{equation}
with 
$\bar{p}_i$ the momentum $p_i$ evaluated at the resonance peak.
The electromagnetic decay width $\Gamma_\gamma$ is given by 
\begin{equation}
\label{a12a32width1} 
\Gamma_{\gamma}=
\frac{\bar p_i^2}
{\pi}\frac{2M_{N}}{(2J_{r}+1)M_{r}} \left[ \left | A_{1/2} \right |^{2}+\left |
A_{3/2} \right|^{2} \right].
\end{equation} 
%

\begin{figure*}
\begin{subfigure}[\ $J_r=3/2$, $\sigma_s=0$ MeV]{
\includegraphics[width=0.9\columnwidth]{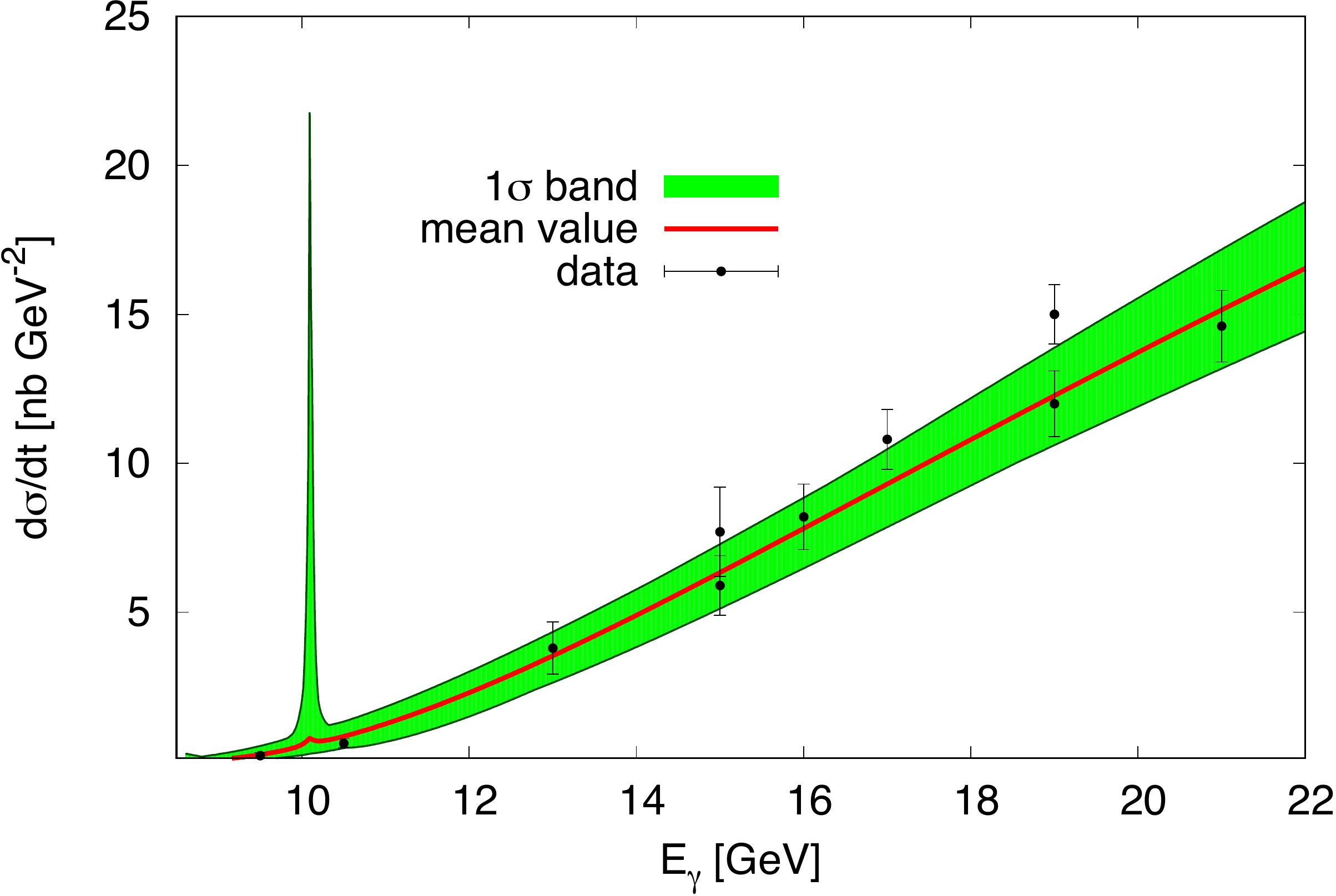}}
\end{subfigure}
\begin{subfigure}[\ $J_r=5/2$, $\sigma_s=0$ MeV]{
\includegraphics[width=0.9\columnwidth]{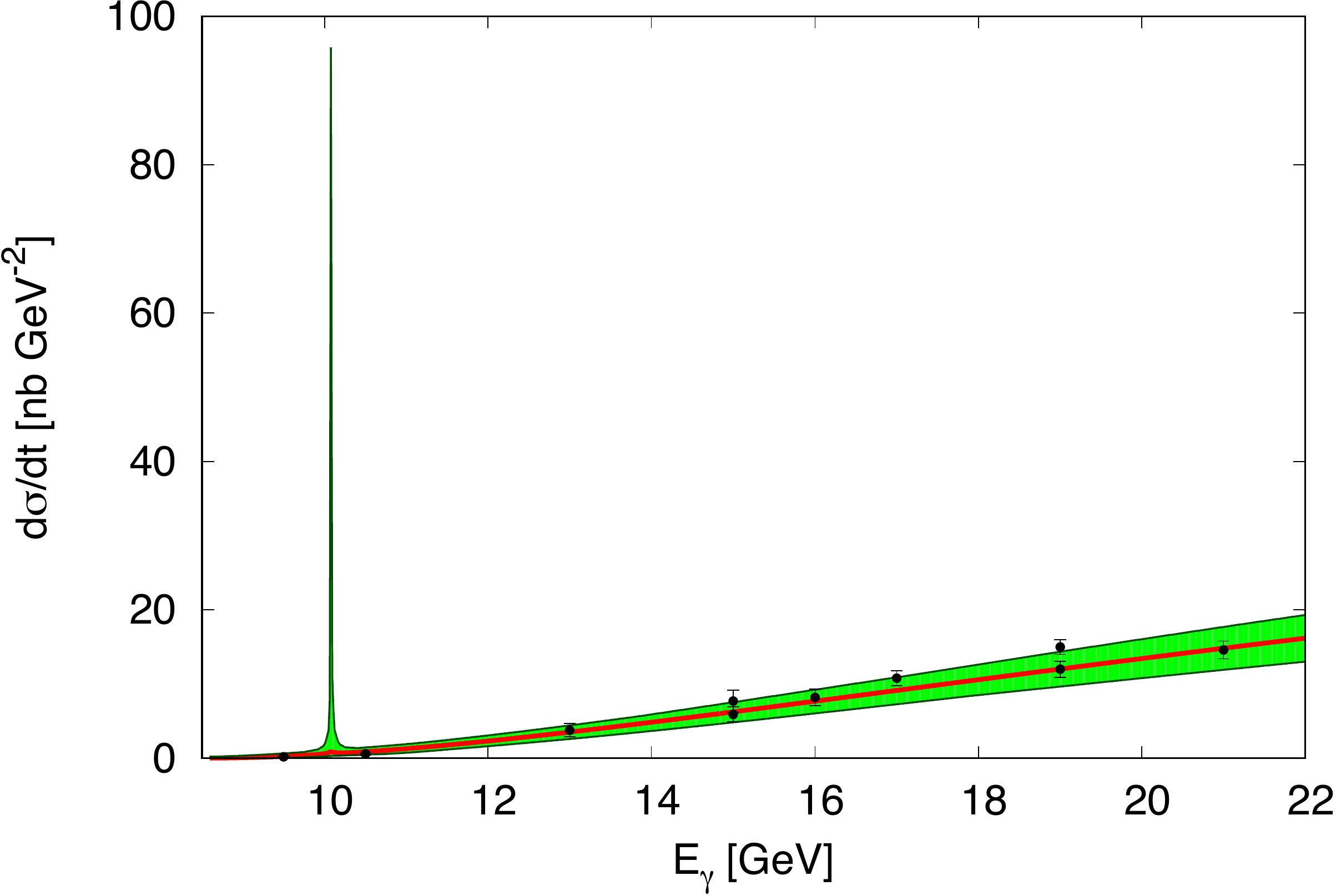}}
\end{subfigure}
\caption{Comparing data (solid circles) with the fit results at a 1$\sigma$ (68\%) CL, 
as discussed in the text, for near-threshold differential cross section 
data~\cite{Ritson:1976rj,Camerini:1975cy} in the forward direction. 
For the cases shown in this figure, no smearing due to experimental resolution is performed.}
\label{FLowLarge}  
\end{figure*}
%

\subsection{Vector meson dominance}
The photon helicity amplitudes for a pentaquark are not known. To rely on data as much as possible, 
we start by following Ref.~\cite{Karliner:2015voa} and assume a VMD 
relation for the transverse vector-meson helicity amplitudes 
\begin{equation}
 \bra{\lambda_{\gamma}\lambda_p}T_\text{em} \ket{\lambda_R} = 
 \frac{\sqrt{4\pi \alpha} f_\psi}{M_\psi} \bra{\lambda_{\psi}=\lambda_\gamma ,\lambda_p}T_\text{dec} \ket{\lambda_R},
 \label{vmd}
\end{equation}
with $f_\psi$ being the \jpsi decay constant which is proportional to the electromagnetic current matrix elements,  
$\bra{0} J^\mu_\text{em} (0)\ket{\jpsi( p,\lambda)} = \sqrt{4\pi \alpha} f_\psi M_\psi 
\epsilon^\mu(p,\lambda).$  The decay constant is related to the $\jpsi$ wave function 
via the Van~Royen-Weisskopf relation, and can be estimated from the leptonic decay width of the $\jpsi \to l^+l^-$, 
yielding  $f_\psi=280~\unit{MeV}$.

Finally, the VMD leads to 
\begin{align}
\Gamma_\gamma=4\pi \alpha \,\Gamma_{\psi p} 
\left(\frac{f_{\psi}}{M_{\psi}}\right)^2 \left(\frac{\bar p_i}{\bar p_f}\right)^{2\ell + 1} \times \frac{4}{6},
\label{eqvecdom}
\end{align}
with the factor $4/6$ due to the fact that in Eq.~\eqref{vmd} only the transverse polarizations of the \jpsi contribute. 
Again, we use $\ell=0$ for $J_r^P=3/2^-$ and $\ell=1$ for $J_r^P=5/2^+$.

With the help of Eqs.~\eqref{a12a32width1} and~\eqref{eqvecdom}, one can constrain the size of the photocouplings.	

\begin{figure*}
\centering
\begin{subfigure}[\ $\sigma_s=0~\unit{MeV}$]{
\includegraphics[width=0.65\columnwidth]{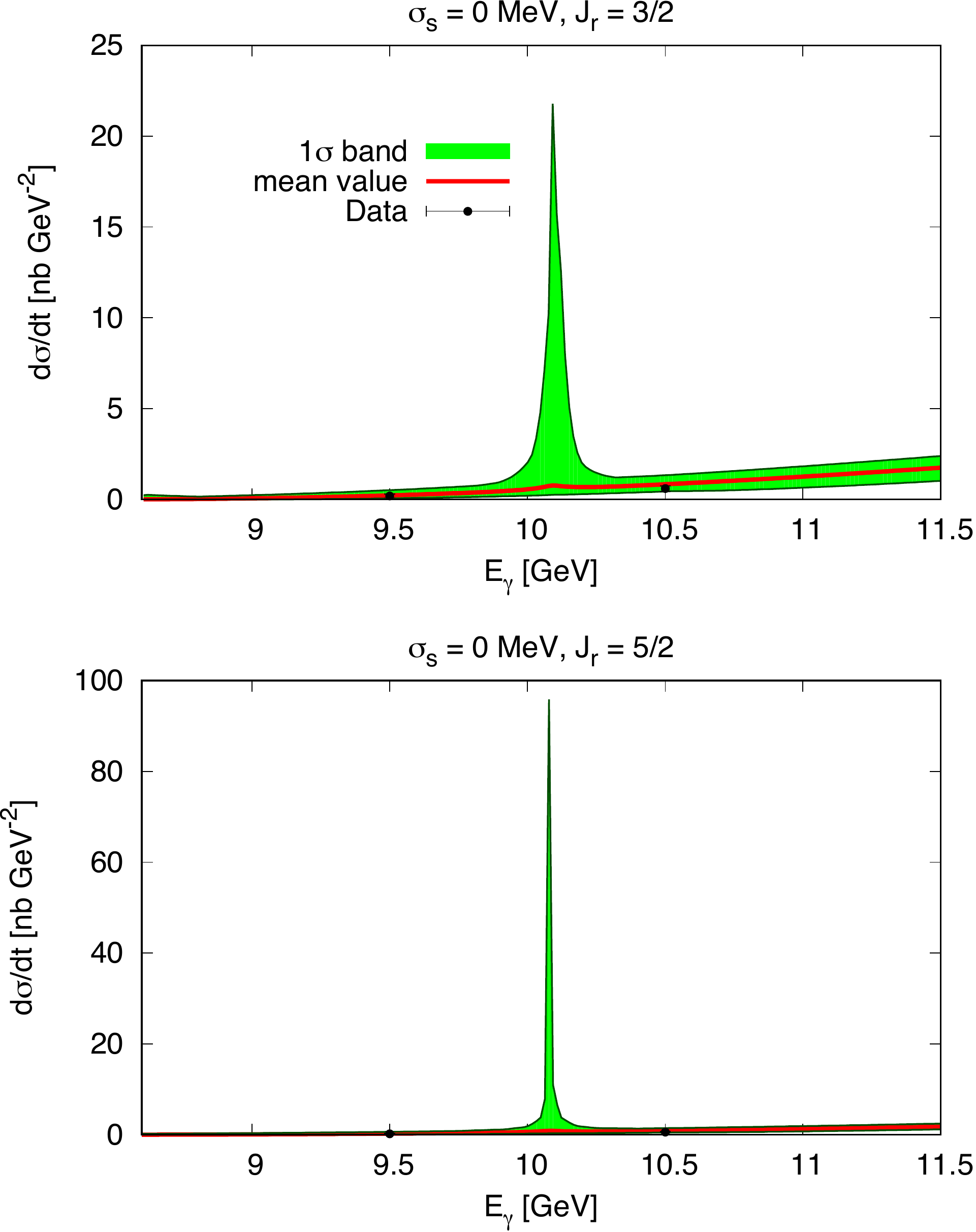}}
\label{FLow}
\end{subfigure}%
\begin{subfigure}[\ $\sigma_s=60~\unit{MeV}$]{
\includegraphics[width=0.65\columnwidth]{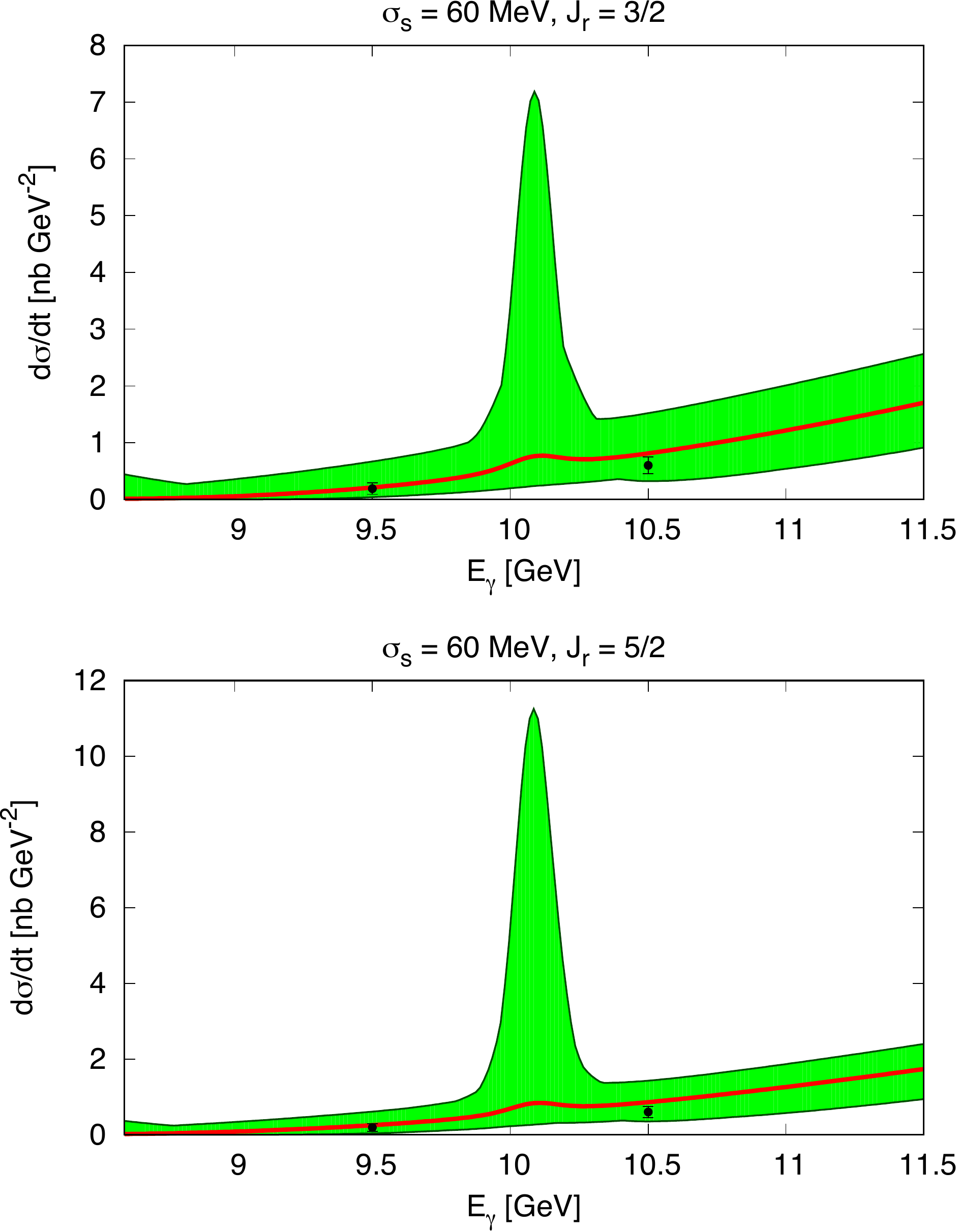}}
\label{FLow60}
\end{subfigure}%
\begin{subfigure}[\ $\sigma_s=120~\unit{MeV}$]{
\includegraphics[width=.65\columnwidth]{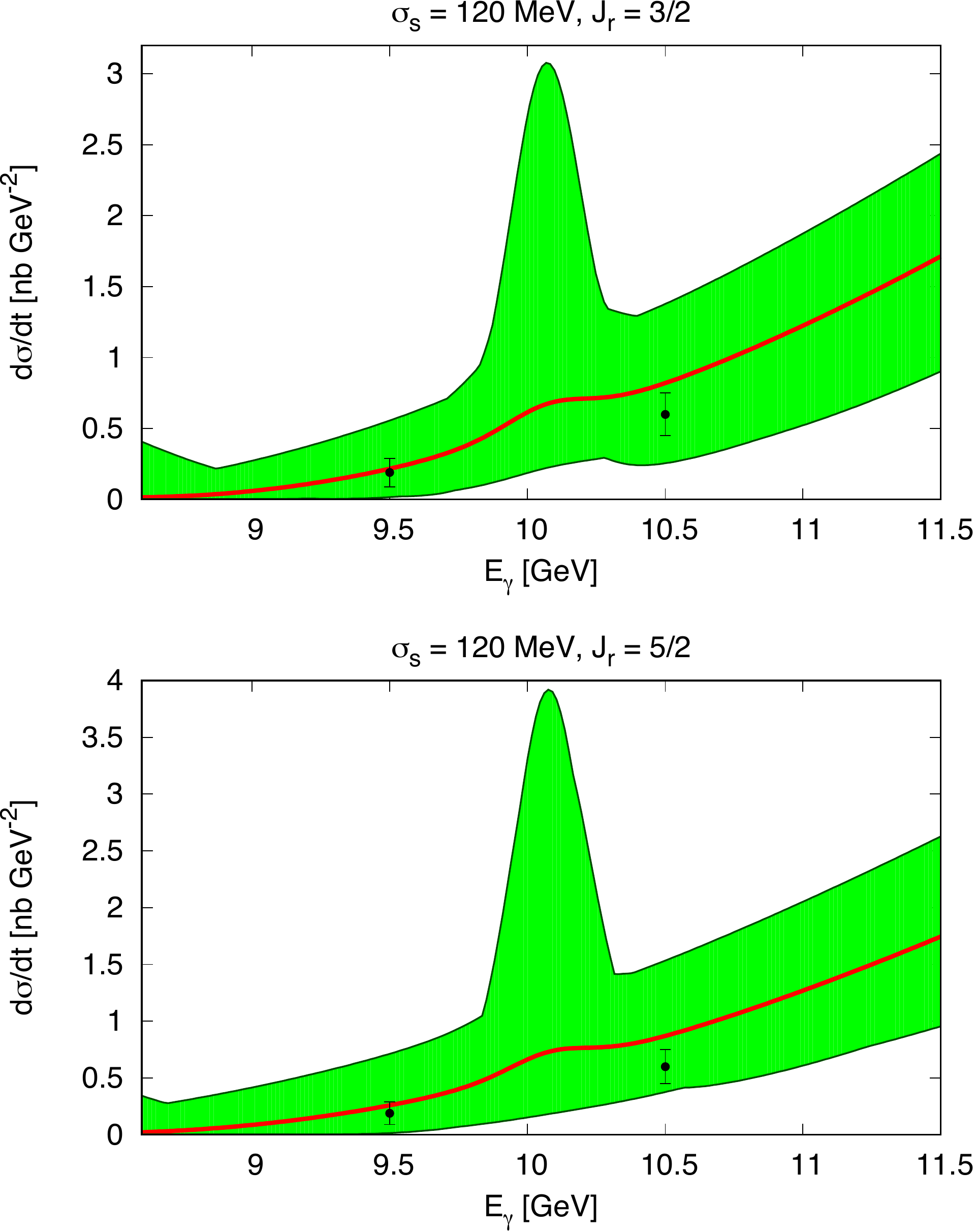}}
\label{FLow120}
\end{subfigure}%
\caption{Comparing data (circles) with the fit results at a 1$\sigma$ (68\%) CL, 
as discussed in the text, for near-threshold differential cross section 
data~\cite{Ritson:1976rj,Camerini:1975cy} in the forward direction. 
Note that the vertical axes have different scales.}
\label{FLowAll}  
\end{figure*}
%
 \begin{table}[b]
\caption{Parameters of the fits with $P_c(4450)$ incorporated as a spin-$3/2$ resonance. 
Uncertainties in the parameters are provided at a 1$\sigma$ ($68\%$) CL except for the
branching ratio $\mathcal{B}_{\psi p}$, whose upper limits are provided at  a $95\%$  CL.
The first line provides the smearing applied to the three lowest-energy experimental data points
in Fig.~\ref{FLowLarge}.} \label{tab:spin32parameters}
\begin{ruledtabular}
\begin{tabular}{cccc}
 $\sigma_s$ (MeV)& 0 & 60 & 120  \\
 \hline
$A$   & $0.156^{+0.029}_{-0.020}$&0.157$^{+0.039}_{-0.021}$ & $0.157^{+0.037}_{-0.022}$ \\
$\alpha_0$   &  $1.151^{+0.018}_{-0.020}$&$1.150^{+0.018}_{-0.026}$ & $1.150^{+0.015}_{-0.023}$\\
$\alpha'$  (GeV$^{-2}$) &  $0.112^{+0.033}_{-0.054}$&$0.111^{+0.037}_{-0.064}$ & $0.111^{+0.038}_{-0.054}$\\
$s_t$  (GeV$^2$)  &  $16.8^{+1.7}_{-0.9}$&$16.9^{+2.0}_{-1.6}$ & $16.9^{+2.0}_{-1.1}$\\
$b_0$ (GeV$^{-2}$)  &  $1.01^{+0.47}_{-0.29}$&$1.02^{+0.61}_{-0.32}$ & $1.03^{+0.49}_{-0.31}$\\
$\mathcal{B}_{\psi p}$   (95\% CL)& $\le 29\: \%$ &$ \le 30\: \%$ & $ \le 23\: \%$\\
\end{tabular}
\end{ruledtabular}
\end{table}
%

\begin{table}[b]
\caption{Parameters of the fits with $P_c(4450)$ incorporated as a spin-$5/2$ resonance. 
Uncertainties in the parameters are provided at a 1$\sigma$ ($68\%$) CL except for the
branching ratio $\mathcal{B}_{\psi p}$, whose upper limits are provided at  a $95\%$  CL.
The first line provides the smearing applied to the three lowest-energy experimental data points
in Fig.~\ref{FLowLarge}.} \label{tab:spin52parameters}
\begin{ruledtabular}
\begin{tabular}{cccc}
 $\sigma_s$ (MeV)& 0 & 60 & 120  \\
 \hline
$A$   & $0.152^{+0.032}_{-0.024}$&$0.150^{+0.043}_{-0.034}$ & $0.150^{+0.044}_{-0.041}$ \\
$\alpha_0$   &  $1.154^{+0.020}_{-0.020}$&$1.156^{+0.027}_{-0.028}$ & $1.156^{+0.033}_{-0.028}$\\
$\alpha'$  (GeV$^{-2}$) &  $0.120^{+0.064}_{-0.052}$&$0.125^{+0.076}_{-0.089}$ & $0.126^{+0.077}_{-0.105}$\\
$s_t$ (GeV$^2$)   &  $16.6^{+1.6}_{-1.1}$&$16.6^{+2.2}_{-1.5}$ & $16.6^{+2.1}_{-2.0}$\\
$b_0$ (GeV$^{-2}$)  &  $0.95^{+0.51}_{-0.51}$&$0.90^{+0.85}_{-0.65}$ & $0.90^{+1.00}_{-0.69}$\\
$\mathcal{B}_{\psi p}$ (95\% CL)  & $\le 17\: \%$ &$ \le 12\: \%$ & $ \le 8\: \%$ \\
\end{tabular}
\end{ruledtabular}
\end{table}
%
\subsection{Nonresonant contribution}
The background in the resonance region
is assumed to be dominated by diffractive production, 
which we parametrize by an effective, helicity-conserving, Pomeron exchange model~\cite{Gribov}, 
\begin{multline}
\langle \lambda_\psi\lambda_{p^\prime}|T_{P} |\lambda_\gamma \lambda_p\rangle = \\
iA~\left(\frac{s-s_t}{s_0}\right)^{\alpha(t)}~e^{b_0(t-t_\text{min})}
\delta_{\lambda_p\lambda_{p'}}\delta_{\lambda_{\psi}\lambda_\gamma}.
\label{EqPom}
\end{multline}
Here $s_0=1\mbox{ GeV}^2$ is fixed. 
Frequently $s_0$ is chosen to match the average $s$ of an  
experiment and that leads to different values for the slope parameter. 
This is unphysical. The physical value of $s_0$ is determined by the range of interactions in the $s$-channel, 
which should be of the order of the hadronic scale.
The Pomeron trajectory is given 
by $\alpha(t)=\alpha_0 + \alpha' \: t $, where $\alpha_0$ and $\alpha'$ 
are parameters to be determined, as well as the normalization $A$, 
the effective threshold parameter $s_t$, and the $t$-slope parameter $b_0$.

There seems to be a rapid decrease of the cross section in the threshold region and the shift parameter  
$s_t$ is introduced to enable a smooth connection between the high energy,  
$W \sim \mathcal{O}(100\mbox{ GeV})$, and the threshold.

%
\section{Results}\label{results}
To the best of our  knowledge, there is no estimate for the
upper limit of the branching ratio of the 
$P_c\rightarrow \jpsi~p$ decay.  
To do so, we aim to fit available data on differential cross sections $d\sigma(\gamma p \to \jpsip)/dt$
with our model given by the coherent sum of the two amplitudes: $T_P$ for the
nonresonant Pomeron background and $T_r$ for the  resonance.

\begin{table*}
\caption{Upper limits at the 2$\sigma$ ($95\%$) CL for the 
resonance parameters obtained for the two  $J_r^P$ assignments. 
We consider three possible values for the experimental resolution $\sigma_s$. 
The resulting values for $g$ follow from 
Eq.~\eqref{ampwidth} and the photocouplings $A_{1/2}$ and $A_{3/2}$ 
are determined using VMD and assuming the couplings 
to be of equal size for the different helicity combinations.} 
\label{tabResParams}
\begin{ruledtabular}
\begin{tabular}{c|ccc|ccc|}
$J_r^P$&\multicolumn{3}{c|}{$3/2^-$}&\multicolumn{3}{c|}{$5/2^+$}\\ \hline
$\sigma_s$ (MeV) &0&60& 120 &0 &60&120 \\
$\mathcal{B}_{\psi p}$ &$ \le 29\%$&$\le 30\%$ &$\le 23\%$&$\le 17\%$&$\le 12\%$&$ \le 8\%$\\
$g~(\unit{GeV})$& $ \le 2.1$&$\le 2.2$&$\le 1.9$ &$\le 2.0$&$\le 1.5$ &$\le 1.4$ \\ 
$\Gamma_{\gamma}~(\unit{keV})$&$\le 14.4$&$\le14.9$&$\le11.0$&$\le 56.9$&$\le 33.5$&$\le 26.8$ \\
$A_{1/2,3/2}~(\unit{GeV}^{-1/2})$ &$\le 0.007$  &$\le 0.007$ &$\le 0.006$&$\le 0.017$ & $\le 0.013$&$\le 0.012$ \\
$\frac{\mathrm{d}\sigma}{\mathrm{d}t}|_{E_\gamma=E_r,t=t_\text{min}}~(\unit{nb}~\unit{GeV}^{-2})$ 
&$\le 21.8$  &$\le 7.2$ &$\le 3.1$&$\le 95.8$ & $\le 11.3$&$\le 3.9$  \\
$\sigma_\text{tot}|_{E_\gamma=E_r}~(\unit{nb})$ &$\le 120$  &$\le 38$ &$\le 14$&$\le 396$ & $\le 44$&$\le 14$  \end{tabular}
\end{ruledtabular}
\end{table*}
The most recent and accurate data for this reaction come from the ZEUS~\cite{Chekanov:2002xi} 
and H1~\cite{Aktas:2005xu} experiments, with a photon energy in the laboratory frame 
$E_\gamma \gtrsim 200~\text{GeV}$. We use the data points with $|t| \le 1.5~\text{GeV}^2$ 
to constrain the Pomeron parameters. For the low-energy region, 
we use data from Camerini~\etal~\cite{Camerini:1975cy}, 
collected at SLAC, that cover the $E_\gamma \sim 13-22~\text{GeV}$ 
energy range. 
To further constrain the fit, we consider the two lowest-energy data points 
shown in SLAC preprints~\cite{Ritson:1976rj}, right  across the pentaquark peak.
In total, we consider 137 $\gamma p \to \jpsip$ data points for the 
$d\sigma(\gamma p \to \jpsip)/dt$ differential cross sections,
with $|t| \le 1.5$ GeV$^2$, covering energy ranges from threshold to 
$E_\gamma \sim 120~\text{TeV}$ in the lab frame.

To compare with the data it is necessary to consider effects of experimental resolution, 
mainly due to the uncertainty in photon energy. 
We introduce a smearing in the calculation of the observables, 
by convoluting the theoretical cross section with a Gaussian distribution
\begin{equation}
G(x)=\frac{\exp \left(-x^2/2\sigma_s^2 \right)}{\sqrt{2\pi}\sigma_s} \, ,
\end{equation}
where $\sigma_s$ is the smearing. The convolution is then given by introducing this function into
\begin{align}
\frac{\mathrm{d}\sigma}{\mathrm{d}\Omega}  \to 
\left(\frac{\mathrm{d}\sigma}{\mathrm{d}\Omega}\ast G\right)(E_\gamma)
=\int_{-\infty}^{\infty}\frac{\mathrm{d}\sigma}{\mathrm{d}\Omega}(y)\,G(E_\gamma-y)\mathrm{d}y,
\end{align}
where $E_\gamma$ is the photon energy in the laboratory frame. 
All the parameters of the model are treated as free parameters in our fits, \ie,
the $P_c(4450)\to \jpsip$ branching ratio $\mathcal{B}_{\psi p}$ and
the Pomeron parameters: $A$, $\alpha_0$, $\alpha'$, $s_t$ and $b_0$.
Only the three lowest-energy data with $t=t_{\text{min}}$ (see Fig.~\ref{FLowLarge}) 
have been smeared with the experimental
resolution.

\begin{figure*}
\centering
\begin{subfigure}[\ Experimental data points from~\cite{Chekanov:2002xi}]{\centering
\includegraphics[width=.65\columnwidth]{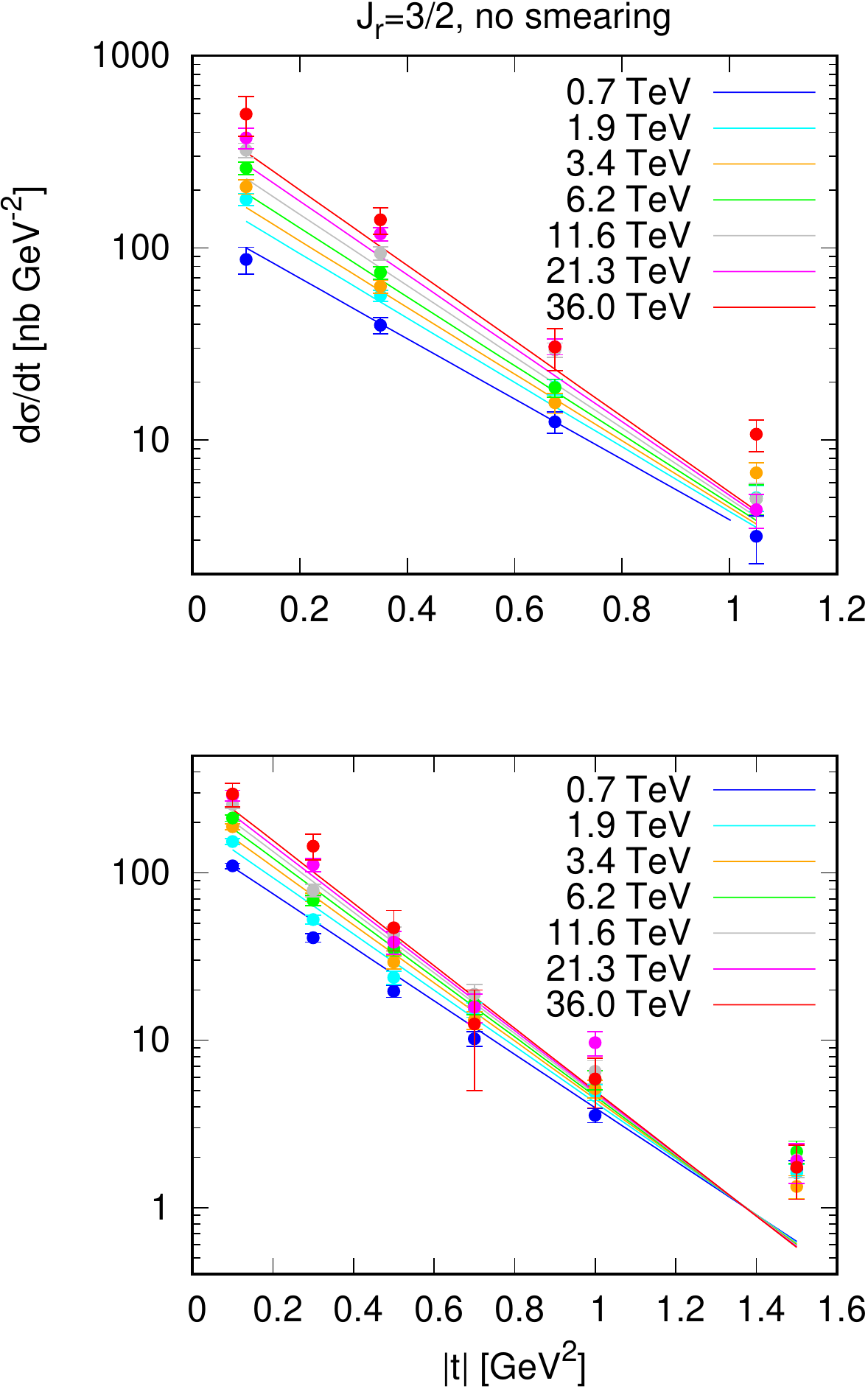}}
\label{FHighOld}
\end{subfigure}\hspace{1.5cm}
\begin{subfigure}[\ Experimental data points from~\cite{Aktas:2005xu}]{\centering
\includegraphics[width=.65\columnwidth]{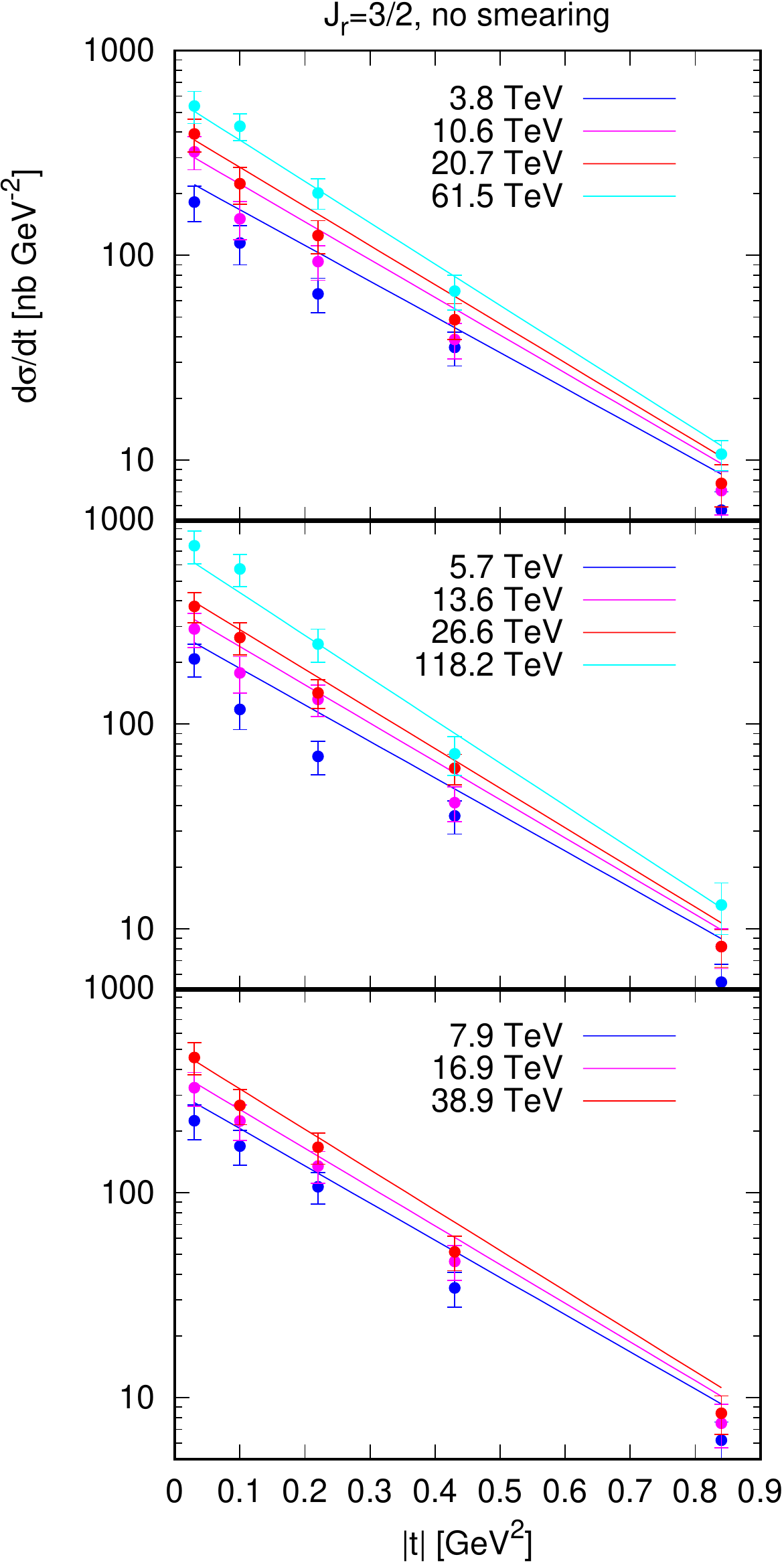}}
\label{FHighNew}
\end{subfigure}%
\caption{Comparing data (circles) with the fit-results' mean values, for $J_r^P = 3/2^-$, and $\sigma_s = 0~\text{MeV}$, and different values of incoming photon energy in the lab frame. 
The error bands are not shown here to ease the plot reading.}
\label{FHighAll}  
\end{figure*}

The mean value of the parameters and the uncertainties have been computed 
employing the bootstrap technique~\cite{bootstrap}.
The bootstrap technique allows to take into account the correlations among parameters 
and to properly propagate their uncertainties to the observables \cite{FR:2016}.
The procedure is as follows.
First we explored the parameter space with $10^5$ fits using \textsc{minuit}~\cite{MINUIT}
in order to identify the region where the absolute minimum lies.
The staring values of the parameters were randomly selected in a very wide range.
Once the parameter-space region where the absolute minimum lies has been identified, 
we use this information 
to randomly seed the starting values of the parameters for our fits. 
We generate $10^4$ data sets for each one of the two $J_r^P$ options ($3/2^-$ and $5/2^+$)
and three smearings ($\sigma_s=0, 60$, and $120$ MeV)
by randomly sampling the experimental points and the pentaquark mass and width according to their uncertainties.
The mass of the $P_c(4450)$ ($M_r=4449.8  \pm 3.2$ MeV) is sampled according to a Gaussian distribution, while
the width, in order to avoid negative values, is sampled according to a Gamma distribution
\begin{equation}
H(x ~|~ \Gamma_r, \sigma_r) = \left( \frac{x\:  \Gamma_r}{\sigma_r^2} \right)^{\frac{\Gamma_r^2}{\sigma_r^2}} 
\frac{\exp \left(-x\: \Gamma_r/\sigma_r^2\right)}{x \: \Gamma(\Gamma_r^2/\sigma_r^2)}\: ,
\end{equation}
where $\Gamma_r=39$ MeV is the pentaquark width and $\sigma_r=24$ MeV its uncertainty.
The experimental data point with the lowest photon energy and $t=t_{\text{min}}$ 
(see Figs.~\ref{FLowLarge} and \ref{FLowAll})
is also sampled according to a Gamma distribution to avoid a negative value of the differential cross section,
while the rest of the experimental data points are sampled according to Gaussian distributions.
For each data set, an independent Maximum Likelihood Fit is performed using \textsc{minuit}~\cite{MINUIT}.
For each smearing option, 
once the $10^4$ data sets have been fitted, we can extract the best $68\%$ fits, 
whose mean value for each parameter provides the best values reported in 
Tables~\ref{tab:spin32parameters} and \ref{tab:spin52parameters}
and the upper and lower values for each parameter provide the uncertainties.
With this set of parameters that correspond to $1\sigma$ confidence level (CL)
we can compute each observable and its uncertainty~\cite{bootstrap}.
Computing quantities at a any other CL (\textit{e.g.} $2\sigma$) can be achieved in the same way,
given that enough fits are computed.
Figure~\ref{FLowAll} shows the
results of the fits for the different smearing parameters
We also show the high-energy data compared to our fit result for the case 
of no smearing in Fig.~\ref{FHighAll}.

Since there are no data for the resonance peak and there is no known lower limit for the branching fraction, the mean values of the parameters are naturally consistent with the non existence of the $P_c(4450)$. We find that the upper limit for the branching ratio $\mathcal{B}_{\psi p}$ at a 95\% CL
ranges from $23\%$ to $30\%$ for $J_r=3/2$, 
depending on the experimental resolution, and from $8\%$ to $17\%$ for $J_r=5/2$. 
The resulting hadronic couplings, as well as the photocouplings, are summarized in Table~\ref{tabResParams}. 
It is worth noting that $A$ is highly correlated with $\alpha_0$,
$\alpha'$ and $b_0$ are also highly correlated, and that
$s_t$ is correlated with $A$ and $b_0$.
$\mathcal{B}_{\psi p}$ is equally correlated with all the parameters.

%
\begin{figure*}[t]
\centering
\includegraphics[width=.78\textwidth]{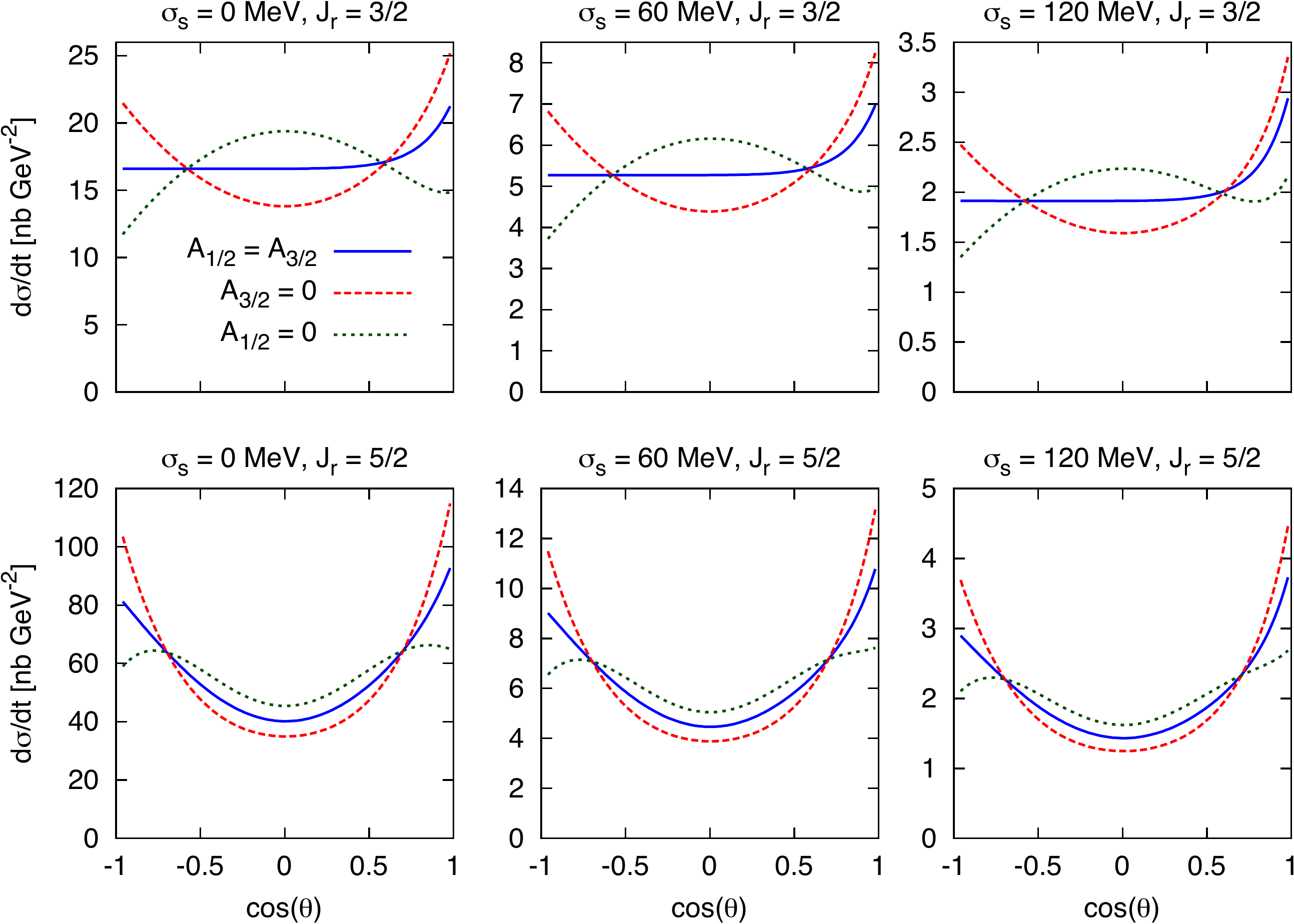}
\caption{Theoretical curves for the differential 
cross section angular distribution at the resonance energy. 
We present the results for the 95\% CL
upper limits for the  branching ratio values, as shown 
in Tables~\ref{tab:spin32parameters} and \ref{tab:spin52parameters} 
for two $P_c(4450)$ spin possibilities and three different photocoupling settings: 
$A_{1/2}=A_{3/2}$ (solid blue);
$A_{3/2}=0$ (dashed red);
$A_{1/2}=0$ (dotted green).}
\label{Fdiffcostheo}
\vspace{1cm}
%

\includegraphics[width=.6\textwidth]{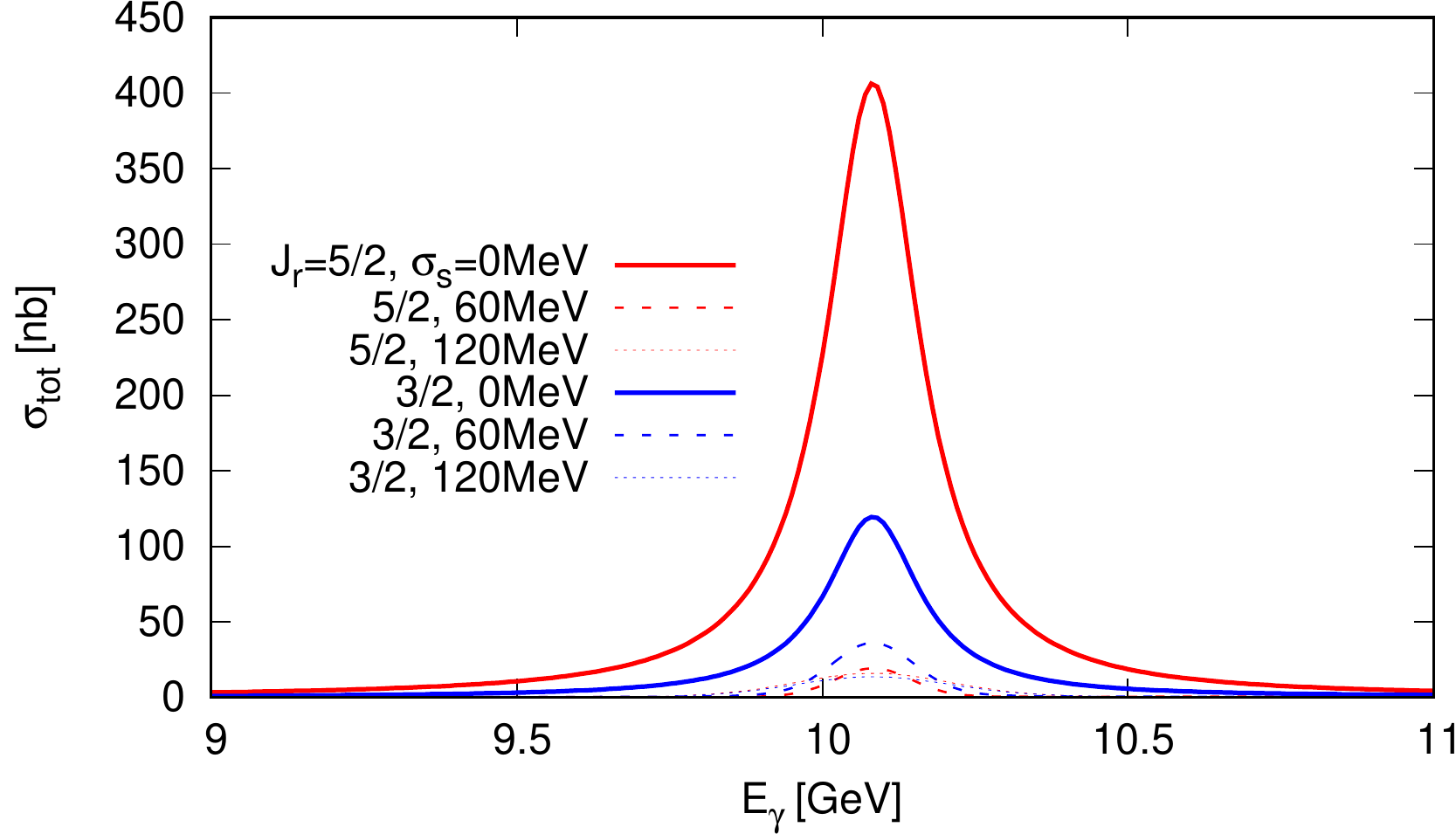}
\caption{The expected total $J/\psi$ photoproduction cross section in the $P_c(4450)$ resonance region, as a function of the lab-frame photon 
energy $E_\gamma$. The chosen photocouplings are  $A_{3/2}=A_{1/2}$ and it is evaluated at a set of fitting parameters where the branching ratio reaches its upper limit, see Tables~\ref{tab:spin32parameters} and \ref{tab:spin52parameters}.}
\label{Ftot}
\end{figure*}

It is possible that a structure at $10~\unit{GeV}$ had indeed not been seen
in the past, because of the finite resolution as the peak strength decreases and becomes broader 
and flatter. 
Its detection requires finer energy  binning. 
The peak is expected 
to be narrow, falling in between two bins where data is available. 
Therefore, it is important to 
perform an energy scan in this region  to confirm the existence of the resonance.

While there are no data on the angular dependence near threshold yet, 
we can make a few predictions based on the model presented above. 
In order to be able to do a deeper study of this dependence, 
we now relax the VMD condition
and let the relative sizes of the photocouplings vary. 
As expected, the results depend on the relative size of the helicity amplitudes, 
as shown in Fig.~\ref{Fdiffcostheo} for three different cases 
$A_{1/2}=A_{3/2}$, $A_{3/2}=0$, and $A_{1/2}=0$, 
while maintaining the sum of the squares $|A_{1/2}|^2+|A_{3/2}|^2$ constant. Pronounced differences in the predicted angular distributions under different assignments for the $P_c(4450)$ photocouplings and spins demonstrate the prospect of determining these quantities from the future experimental data on $J/\psi$ photoproduction cross sections measured with CLAS12 in the near-threshold region.

{Finally, we discuss the results for the total cross section, 
shown in Fig.~\ref{Ftot} for the two possible spin assignments $3/2$ and $5/2$, 
and the different experimental resolutions shown in Tabs.~\ref{tab:spin32parameters} to \ref{tabResParams}. 
The resonant piece of the resulting cross section is consistent with the one predicted by Ref.~\cite{Karliner:2015voa} 
when taking the same spin and branching-ratio assumptions: $J_r=3/2$ and $\mathcal{B}_{\psi p}=10\%$.} 

\section{Summary}
We studied the possibility of observing the $P_c(4450)$ resonance in $\jpsi$ photoproduction at CLAS12. 
The data available in this energy range are scarce and there is no measurement of the differential cross section. 
We tested the compatibility of the data and a simple two-component model 
containing the directly produced resonance and a diffractive background. 
We conclude that, the resonance peak being narrow enough, it could have escaped detection due to poor energy resolution.

From the fits to the available $\jpsi$ photoproduction data,
we show that the magnitude of the peak for the $P_c(4450)$ resonance 
can range from very strong to barely visible. 
Our results demonstrate the possibility to observe 
the $P_c(4450)$ resonance and to determine its spin and photocouplings. 
Therefore, the future data on $\jpsi$ photoproduction off protons measured 
in the near threshold region with a quasi-real photon beam at JLab will allow us to explore 
photoexcitation of the $P_c(4450)$ state suggested by LHCb data.
 
We made a prediction for the angular distribution which can be used to 
disentangle the spin of the resonance and helicity dependence of the resonance production. 
If the resonance  signal is found, photon-virtuality dependence can be used to investigate the resonance structure. 

\begin{acknowledgments}
This material is based upon work supported in part by the 
U.S.~Department of Energy, 
Office of Science, 
Office of Nuclear Physics under contracts
DE-AC05-06OR23177 and DE-FG0287ER40365, 
National Science Foundation under grants PHY-1415459 and NSF-PHY-1205019, and IU Collaborative Research Grant. 
This work was also supported by the Spanish Ministerio de Econom\'{\i}a y Competitividad (MINECO)
and European FEDER funds under contract No.~FIS2014-51948-C2-2-P and SEV-2014-0398. 
ANHB acknowledges support from the Santiago Grisol\'{\i}a program 
of the  Generalitat Valenciana and from the Center for Exploration of Energy and Matter at Indiana University. 
\end{acknowledgments}


\end{document}